\documentclass[]{spie}  

\usepackage{amsmath,amsfonts,amssymb}
\usepackage{graphicx}
\usepackage[colorlinks=true, allcolors=blue]{hyperref}
\usepackage{amssymb,amsmath}
\usepackage{wrapfig}

\title{MagAO-X: project status and first laboratory results}

\author[a]{Jared R. Males}
\author[a]{Laird M. Close}
\author[a,b]{Kelsey Miller}
\author[a,b]{Lauren Schatz}
\author[c]{David Doelman}
\author[a,b]{Jennifer Lumbres}
\author[c]{Frans Snik}
\author[a,b]{Alex Rodack}
\author[a,b]{Justin Knight}
\author[a,b]{Kyle Van Gorkom}
\author[a]{Joseph D. Long}
\author[a,b]{Alex Hedglen}
\author[a,b]{Maggie Kautz}
\author[d]{Nemanja Jovanovic}
\author[a]{Katie Morzinski}
\author[a,b,e,f]{Olivier Guyon}
\author[g]{Ewan Douglas}
\author[h]{Katherine B. Follette}
\author[e]{Julien Lozi}
\author[a]{Chris Bohlman} 
\author[a]{Olivier Durney}
\author[a]{Victor Gasho}
\author[a]{Phil Hinz}
\author[i]{Michael Ireland}
\author[a,b]{Madison Jean}
\author[c]{Christoph Keller}
\author[c]{Matt Kenworthy}
\author[j]{Ben Mazin}
\author[a]{Jamison Noenickx}
\author[a]{Dan Alfred}
\author[a]{Kevin Perez}
\author[a]{Anna Sanchez}
\author[a]{Corwynn Sauve}
\author[k]{Alycia Weinberger}
\author[l]{Al Conrad}

\affil[a]{Steward Observatory, University of Arizona} \affil[b]{College of Optical Sciences, University of Arizona}
\affil[c]{Sterrewacht Leiden, Universiteit Leiden}
\affil[d]{Caltech Optical Observatory, California Institute of Technology}
\affil[e]{Subaru Telescope, National Astronomical Observatory of Japan}
\affil[f]{Astrobiology Center, National Institutes of Natural Sciences, Japan}
\affil[g]{Massachusetts Institute of Technology, Department of Aeronautics and Astronautics}
\affil[h]{Physics and Astronomy Department, Amherst College}
\affil[i]{The Australian National University}
\affil[j]{Department of Physics, University of California Santa Barbara}
\affil[k]{Department of Terrestrial Magnetism, Carnegie Institution for Science}
\affil[l]{Large Binocular Telescope Observatory }

\authorinfo{Further author information: (Send correspondence to JRM)\\JRM: E-mail: jrmales@email.arizona.edu}

\begin{document} 
\maketitle

\begin{abstract}
MagAO-X is an entirely new “extreme” adaptive optics system for the Magellan Clay 6.5 m telescope, funded by the NSF MRI program starting in Sep 2016.  The key science goal of MagAO-X is high-contrast imaging of accreting protoplanets at H$\alpha$.  With 2040 actuators operating at up to 3630 Hz, MagAO-X will deliver high Strehls ($>70$\%), high resolution (19 mas), and high contrast ($< 1\times10^{-4}$) at H$\alpha$ (656 nm).  We present an overview of the MagAO-X system, review the system design, and discuss the current project status.  
\end{abstract}


\section{INTRODUCTION}
\label{sec:intro}  
AO systems are now in routine use at many telescopes in the world; however, nearly all work only in the infrared (IR, $\lambda>1$ $\mu$m) due to the challenges of working at shorter wavelengths.  The Magellan AO (MagAO) system was the first to routinely produce visible-AO science on a large aperture telescope\cite{2013ApJ...774...94C,close_2014_hd142527, 2014ApJ...786...32M, rodigas_2015, sallum_nature_2015} (see Close et al. in these proceedings \cite{laird_magao}).     Other large telescopes with visible AO systems include the 5 m at Palomar \cite{dekany_2013} and ESO's 8 m VLT with the ZIMPOL camera behind the Spectro-Polarimetric High-contrast Exoplanet REsearch (SPHERE) instrument \cite{roelfsema_2014}.

MagAO-X is an \textit{entirely new} visible-to-near-IR ``extreme'' AO (ExAO) system.  When completed, MagAO-X will consist of: (1) a 2040 actuator deformable mirror (DM) controlled at (up to) 3.63 kHz by a pyramid wavefront sensor (PWFS); (2) cutting-edge coronagraphs to block a star's light; and (3) a suite of focal plane instruments including imagers and spectrographs enabling high-contrast and high-resolution science.

MagAO-X will deliver high Strehls ($\gtrsim 70\%$ at H$\alpha$), high resolutions ($14-30$ mas), and high contrasts ($\lesssim 10^{-4}$) from $\sim$ 1 to 10 $\lambda/D$ .  Among many compelling science cases, MagAO-X will revolutionize our understanding of the earliest stages of planet formation, enable high spectral-resolution imaging of stellar surfaces, and could take the first images of an exoplanet in reflected light.

\subsection{Status}

MagAO-X is funded by the NSF MRI program (award \#1625441), beginning in Sep, 2016.  We passed a rigorous external preliminary design review (PDR), which was managed by the Magellan Observatory, in May, 2017.  The detailed design was completed, and procurement is nearly complete.  The last major item to be delivered is the MEMS deformable mirror.  Integration is underway in the Extreme Wavefront Control Lab (EWCL) in Steward Observatory at the University of Arizona.

Our current plan is to hold the Pre-Ship Review, a Magellan Telescope requirement, in early 2019, and then ship to Las Campanas Observatory (LCO) in mid to late Feb, 2019.  Technical first light would then occur in roughly March 2019.  The instrument will be shipped back and forth from LCO to Tucson several times, at least until commissioning is complete in late 2020.

Here we provide a very abbreviated overview of the MagAO-X design.  Reference is made to the many other presentations in this conference providing more detail.  In addition, the complete PDR documentation is available at \url{https://magao-x.org/docs/pdr}.

\section{SCIENCE JUSTIFICATION}

MagAO-X will enable a wide range of astrophysical observations.  Here we review a few of these.

\subsection{A Survey of the Low Mass Distribution of Young Gas Giant Planets} 

The main science goal of the initial operations of MagAO-X is to conduct a survey of nearby T Tauri and Herbig Ae/Be stars for newly formed accreting planets in H$\alpha$.  

\subsubsection{Proof-of-concept -- Imaging LkCa 15 b at H$\alpha$} 

We used MagAO's simultaneous differential imaging (SDI) mode \cite{close_2014_hd142527} to discover H$\alpha$ from a forming protoplanet (LkCa 15 b)\cite{sallum_nature_2015}, detecting an accretion stream shock at 90 mas.  Dynamical stability places the mass of LkCa 15 b at $2^{+3}_{-1.5}$ $M_{Jup}$.  Correcting for extinction we found an accretion rate of $\dot{M} = 1.16\times10^{-9}$ $M_{\odot}/yr$ \cite{sallum_nature_2015}.  Observed $H{\alpha}$ rate was $\sim0.5e/s$ at the peak pixel at Strehl $\sim$5\%.

\subsubsection{MagAO-X $H\alpha$ Protoplanet Survey} 

While LkCa 15 (145 pc; 1--2 Myr) is fairly faint ($I$$\sim$$11$ mag) there are many brighter, closer similarly young accreting targets.  A review of the populations of nearby young moving groups and clusters yields 160 accreting, $<$10 Myr old,  D$<$150 pc, targets all with I$<$10 mag for this survey---bright enough for good AO correction. There are also 33 more stars with I$<$12 mag that will be excellent targets with moderate to good AO correction.  Extrapolating from our initial ($3/10=30\%$) success rate for the young star GAPplanetS $H\alpha$ survey having accreting objects \cite{Fol_2018} the 193 stars yield  $\sim$$59$ new protoplanet systems using just 5 nights per semester.  

This survey of protoplanet systems will define the population of low-mass outer EGPs, and will help reveal where and how gas planets actually form and grow.  As the main initial science case for MagAO-X, these observations define the performance requirements of our design.  In Table \ref{tab:high_reqs} we detail the guide star statistics, and the corresponding system performance requirements needed to enable this survey.

\begin{table}[h!]
\centering
\caption{The high level performance requirements derived from  H$\alpha$ SDI survey. \label{tab:high_reqs}}
\begin{tabular}{ccc||ccc}
\hline
\multicolumn{3}{c||}{Targets} & \multicolumn{3}{c|}{Performance} \\
\hline
I     & d     &  Numb. & Sep  & $\Delta$H$\alpha$  & Strehl$^1$ \\
mag  & [pc]  &         & [mas] & mag              &  [\%]     \\
\hline
5      & 225  & 6 & 75    & 12.0 & 70  \\
8      & 150  & 25  & 100   & 9.0 & 50  \\
10     & 150  & 129  &100   & 7.0 & 30  \\
12     & 150 & 44$^2$  & 100    & 5.0 & 20   \\
\hline
\multicolumn{6}{l}{$^1$ At H$\alpha$, $\lambda=$656 nm.}\\
\multicolumn{6}{l}{$^2$ not complete, we consider this a lower limit}\\
\end{tabular}
\end{table}

\subsection{Other Science Cases}

Here we present a short summary of several additional science cases.    These are generally spanned by the parameters of the H$\alpha$ survey in terms of guide star brightness and separations, and are less-demanding, and so we do not specifically derive requirements for these.  Rather, they are presented to give an idea of the breadth of use-cases for this new instrument.

\subsubsection{Circumstellar Disks}  Disk science is a challenging application of AO, with low surface brightness and characteristics similar to the uncorrected seeing halo, so high-Strehl high-contrast ExAO is critical.  MagAO-X will push the two frontiers in circumstellar disk science.  The first is detailed imaging of geometry, particularly in the 5-50 AU region analogous to the outer part of the solar system. Most disks sit at 50-150 pc, so reaching radii comparable to the giant planet region requires imaging at 50-120 mas. Existing systems push in to at best $\sim$150 mas. For some disks, an inner working angle of $\sim$100~mas will
push to the exozodiacal light region for the first time. For example, in the
well-known HR 4796A disk, SED fits show that the 8-20 $\mu$m flux cannot be fully explained by the outer, $\sim$100~K, ring, suggesting a ring at 3-7 AU \cite{wahhaj_2005}. MagAO-X has the
potential to image this inner ring.  
The second frontier is multi-wavelength study of disks to derive the
chemical make-up and dynamical state \cite{rodigas_2015,stark_2014}.  This requires a large wavelength grasp from visible through near-infrared so MagAO-X's ability to
image at $\sim$0.45 $\mu$m complements existing systems.

\begin{figure}[b!]
\centering
\includegraphics[height=2in]{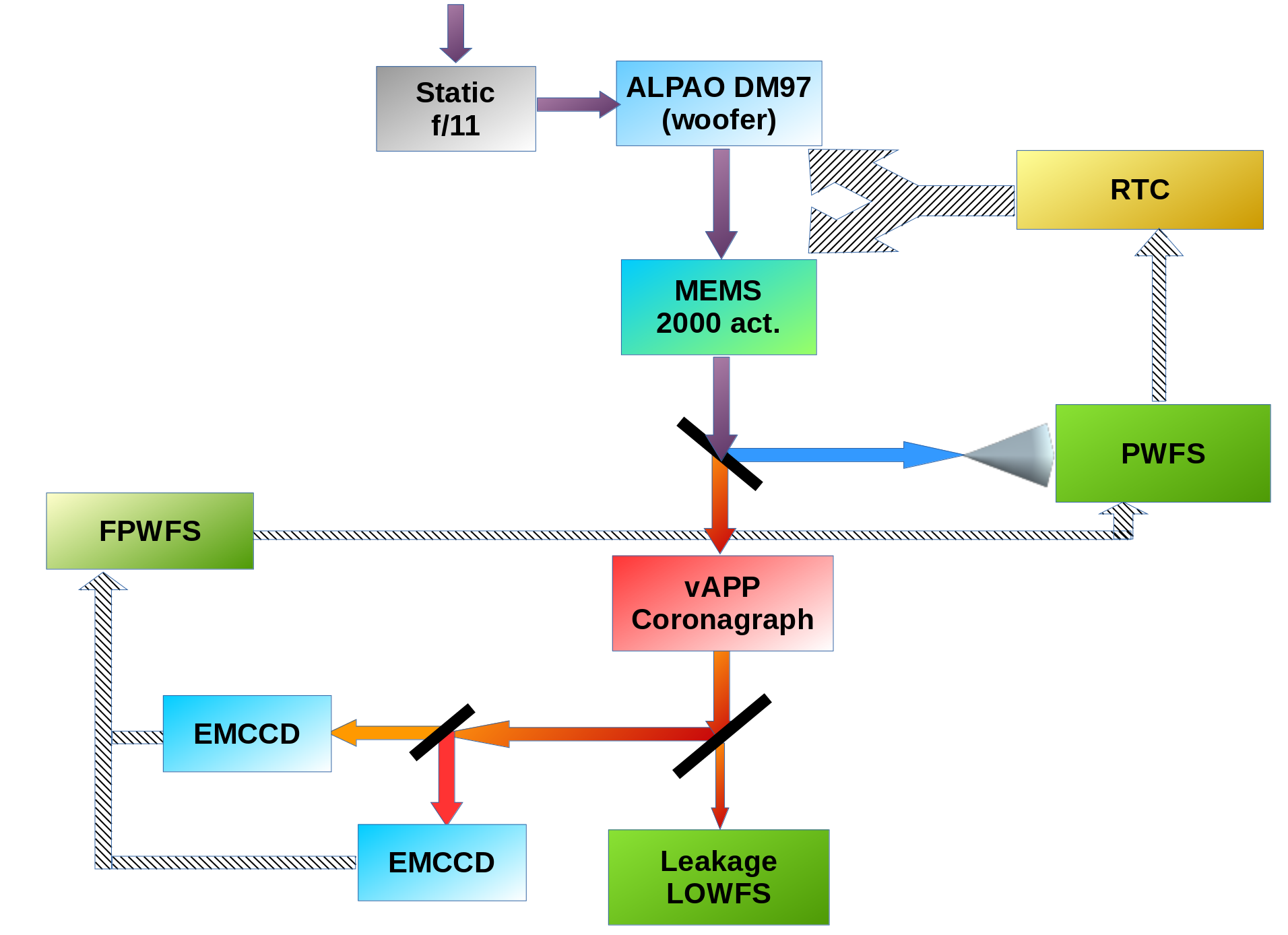}
\includegraphics[height=2in]{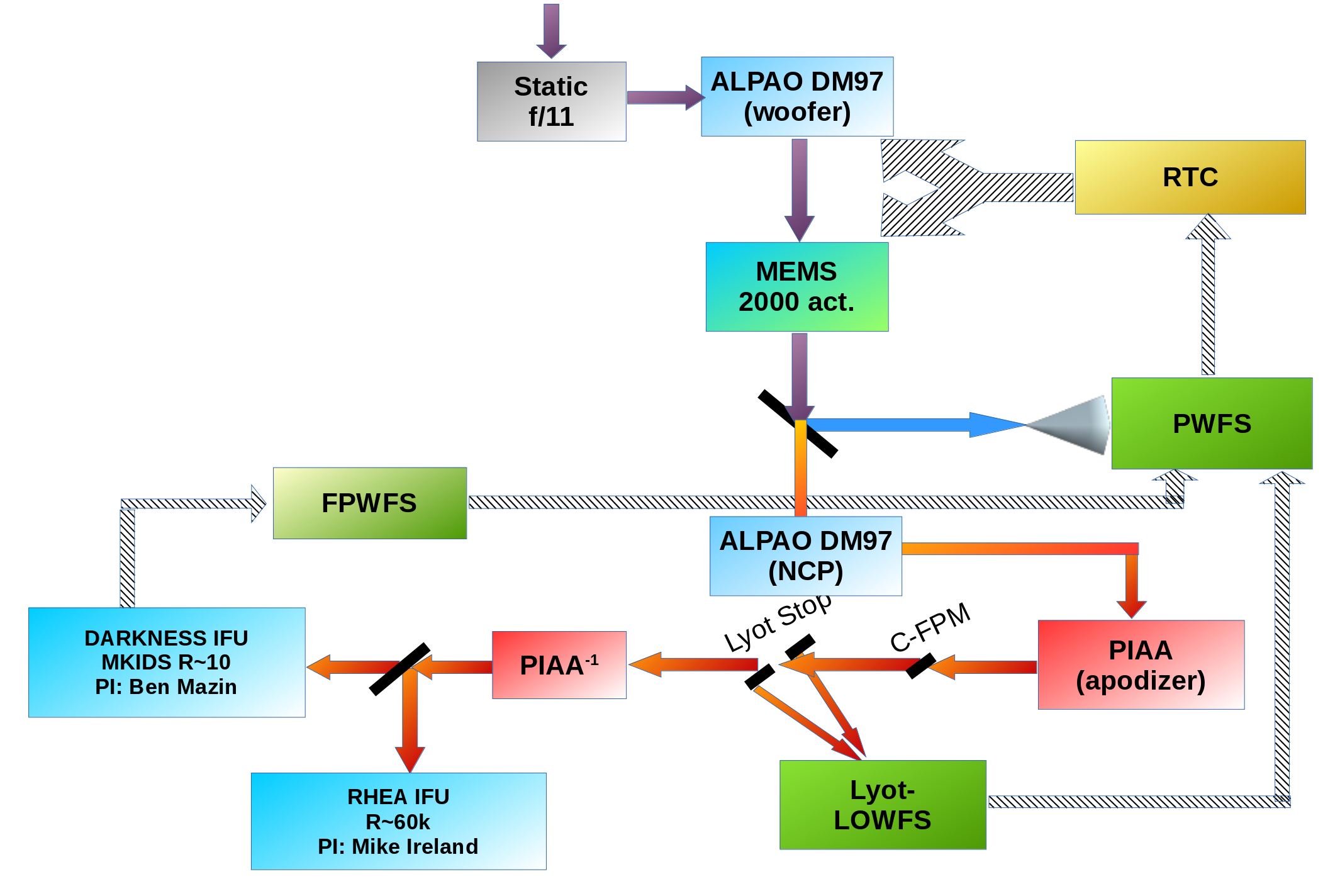}
\caption{\label{fig:wt_block} The MagAO-X woofer-tweeter architecture. Left: in Phase-I we employ the vAPP coronagraph and SDI.  Right: in Phase-II we will employ a PIAACMC and spectrographs.}
\end{figure}

\subsubsection{Fundamental Properties of Young Solar-System-like EGPs}

Dedicated exoplanet-imagers GPI and SPHERE are in operation, and GPI has discovered the first planet of this new era: 51 Eri b is a 600-K $\sim$2 M$_\textrm{Jup}$ exoplanet imaged 13 AU from its 20-Myr-old, 30-pc-away F-type host star \cite{2015Sci...350...64M}.  This planet is different from other exoplanets (whether imaged or analyzed by transit spectroscopy):  its atmosphere is the closest analog yet to solar system atmospheres because of its Saturn-scale orbit, Jupiter-scale mass, and cool temperature such that CH$_4$ was detected in the GPI spectrum.

We have conducted a prototype experiment with existing MagAO using the exoplanet $\beta$ Pic b, which can be imaged with the current VisAO due to its brightness (youth and mass) and its 300-400 mas separation\cite{2014ApJ...786...32M}.  We also demonstrated using such measurements to empirically measure the fundamental properties of this solar-system-scale exoplanet\cite{2015ApJ...815..108M}.   MagAO-X will extend these observations to shorter wavelengths, and to smaller mass, smaller separation planets such as 51 Eri b.  MagAO-X will also enable characterization of such planets with the DARKNESS and RHEA@MagAO-X spectrographs.

\subsubsection{Resolved Stellar Photospheres}

High resolution spectroscopy will be enabled by the 3x3 single-mode fiber-fed IFS, RHEA@MagAO-X, with $R\sim$60,000 spectral resolution (PI: Mike Ireland).  The combination of the ExAO resolution and contrast with high spectral resolution enables many exciting science cases. For instance: the largest resolvable non-Mira stars accessible from MagAO include Betelgeuse ($\sim$50\, mas), Antares ($\sim$40\, mas), Arcturus ($\sim$21\, mas), Aldebaran ($\sim$20\, mas) and  $\alpha$~Boo ($\sim$19\, mas).  These stars lose mass through a complex process in an interplay between a hot ($\sim$10,000 K) corona and a cool ($\sim$2000 K), slow ($\sim$10\, km/s) molecular wind.  These states can not co-exist so asymmetries of some kind are expected.  Resolving the photosphere in lines and molecular bands enables the multi-dimensional structure of these regions to be imaged.  Upwelling and downwelling velocities on the surface are of order a few km/s, separable at sufficient resolution.  A single image of a stellar photosphere would be the first ever direct measurement of convection in a star other than the Sun.

\subsubsection{Asteroids} 

MagAO-X will have resolutions of 14--21 mas in $g$-$r$ bands, which correspond to $\sim$20--30 km on a main-belt asteroid (MBA).  On a typical night more than 80 MBAs brighter than I=13 (implying $\gtrsim50$ mas) will be resolvable by MagAO-X.  This will provide true dimensions, avoiding degeneracies in light-curve analysis.  MagAO-X will enable sensitive searches for and orbit determination of MBA satellites.  In combination, these directly measure density and hence estimate composition \cite{britt_2002}.  This will directly inform the theories of terrestrial planet formation \cite{mordasini_2011}.  See Ref. \citenum{2017Icar..281..388S} for a proof of concept with existing VisAO.

\section{OPTOMECHANICAL DESIGN}

\subsection{System Architecture}

\begin{wrapfigure}[20]{r}{3in}
\vspace{-0.5in}
\includegraphics[width=3in]{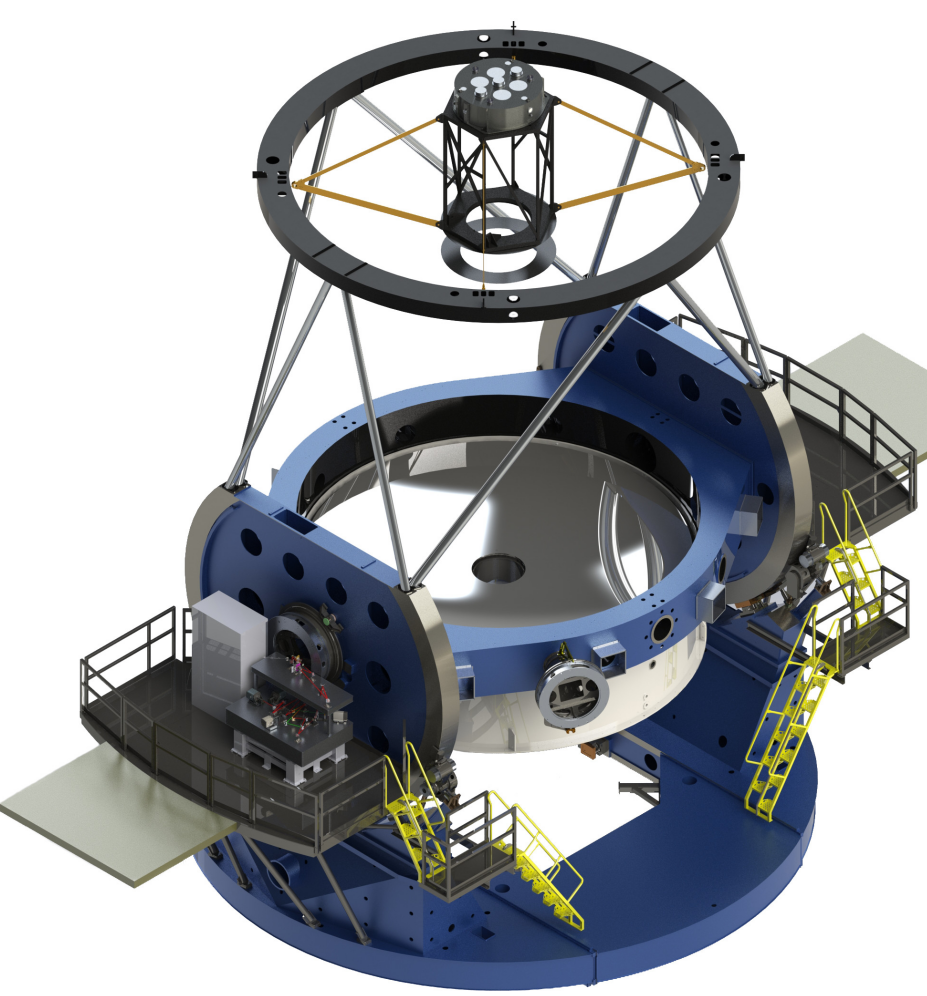}
\caption{\label{fig:magaox_clay} MagAO-X on the 6.5 m Magellan Clay telescope Nasmyth-East platform.  }
\end{wrapfigure}

Our original concept for MagAO-X was to employ a cascaded control system, where the existing MagAO system provided an initial low-order low-speed correction, followed by a high order ``afterburner'' system.  This is similar to the current SCExAO architecture\cite{2015PASP..127..890J,julien_scexao}.  During the preliminary design phase of the project, we opted to prioritize a self-contained woofer-tweeter system, using the static facility f/11 secondary at Clay.  The main driver of this choice is calibration: full calibration and end-to-end testing of the system with the MagAO adaptive secondary mirror would require occupying the actual telescope with a retroreflector installed.  A secondary driver is operational, as using the f/11 (which is normally installed) provides much more flexibility compared to requiring the f/16 ASM to be installed to support MagAO-X operations.  With the self-contained woofer-tweeter architecture, MagAO-X is a completely separate AO system from existing MagAO.

Figure \ref{fig:wt_block} shows the high level architecture of MagAO-X.  Light enters from the f/11 secondary (off the tertiary) and the pupil is relayed to an Alpao DM-97 woofer.  Another pupil image is formed on the Boston Micromachines (BMC) 2k DM.  A beamsplitter then splits light (either 50/50 or with dichroics) between the PWFS and the science channel.  The PWFS signal is reconstructed and fed back to the DMs in closed loop.  The science light is passed through the coronagraph.  At left we show our ``Phase I'' plan, which includes a vector Apodizing Phase Plate (vAPP) coronagraph, followed by a simultaneous differential imaging system which makes use of dichroic beamsplitters and two EMCCD cameras.   Light rejected by the coronagraph, and/or light contained in the vAPP leakage term is used for coronagraph low-order WFS (LOWFS). At right is the ``Phase II'' plan, which will employ a Phase Induced Amplitude Apodization Complex Mask Coronagraph (PIAACMC) with a Lyot-LOWFS\cite{2017PASP..129i5002S}.  In this phase we will also deploy spectrographs including the DARKNESS MKIDS\cite{2018PASP..130f5001M,alex_mec} array-IFU and the RHEA single-mode fiber-fed IFU.

\subsection{Detailed Design}

MagAO-X consists of an air-isolated optical table with an active height control and leveling system.  It has two levels.  The upper level is at the height of the telescope beam above the Nasmyth platform, and contains the woofer, or low-order DM (LODM), the tweeter, or high-order DM (HODM), the image re-rotator (K-mirror), and the atmospheric dispersion corrector (ADC).  The lower level houses the PWFS, various filter wheels and beamsplitter selectors, the coronagraph optics, and the science cameras.

\begin{figure}[t!]
\centering
\includegraphics[height=2.5in]{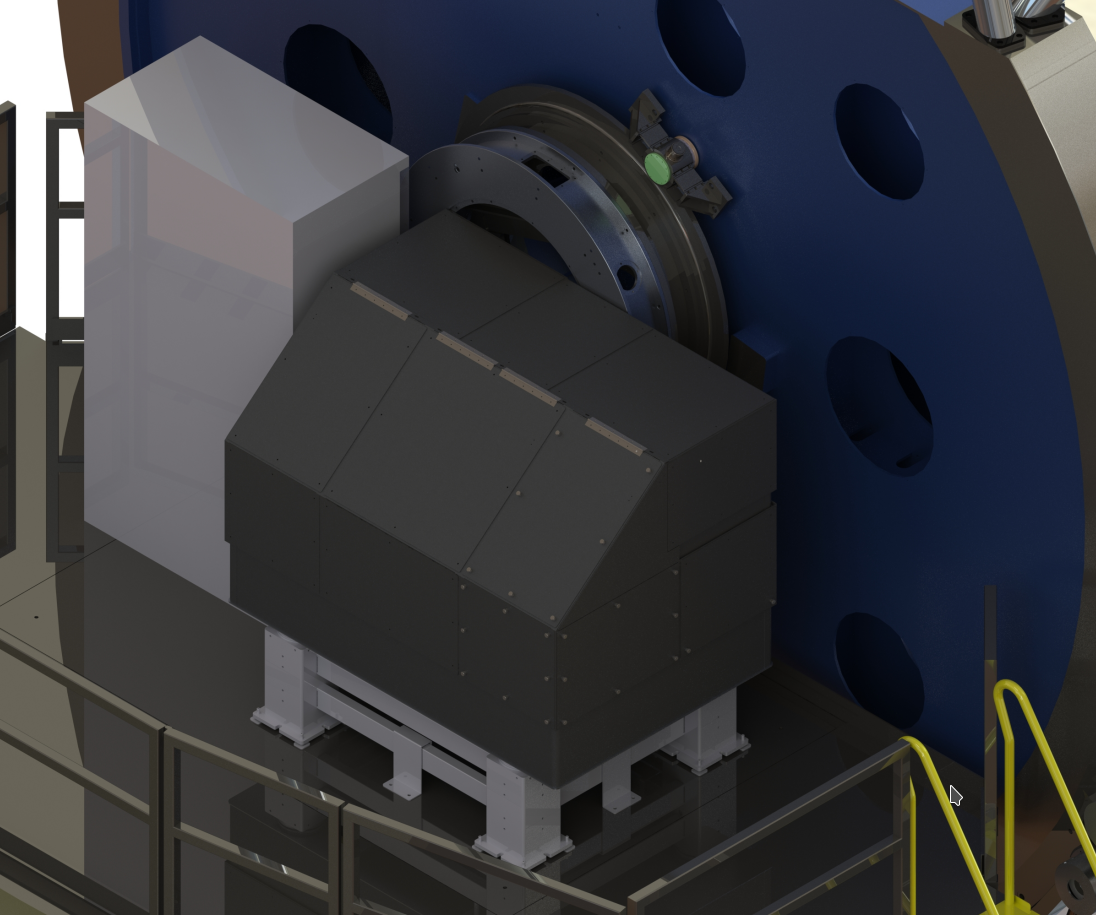}
\includegraphics[height=2.5in]{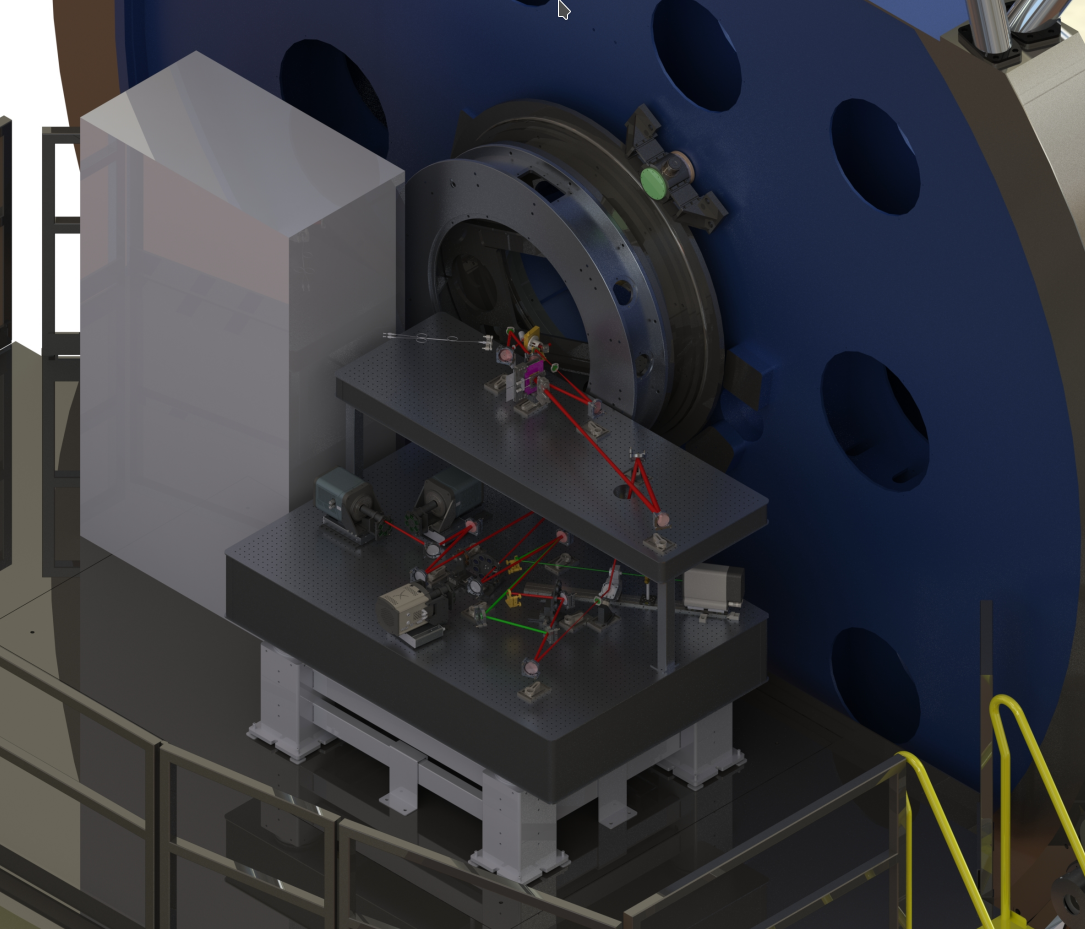}
\caption{\label{fig:magaox_cover} MagAO-X closeup.  Left: the instrument with dust cover installed.  Right: dust cover removed to show the details inside.}
\end{figure}

The design of MagAO-X is illustrated in Figure \ref{fig:magaox_clay}, showing the MagAO-X optical bench on the Nasmyth platform of the Clay telescope, along with the electronics rack.  Figure \ref{fig:magaox_cover} shows a closeup of the instrument.  A dust cover will protect the instrument, with a slight positive pressure to prevent dust contamination.  It is not under vacuum nor is it thermally controlled.  For more information about the optomechanical design of MagAO-X see Close et al. in these proceedings\cite{laird_optomech}.

Since MagAO-X will be deployed on the Nasmyth platform of an Alt-Az telescope, we require an image derotator to keep the pupil aligned to the WFS, DMs, and coronagraphs.  For a description of the custom K-mirror used to accomplish this see Hedglen et al. in these proceedings\cite{alex_k}

Since MagAO-X is not thermally controlled, thermally stable optics mounts are critical to maintaining the fine alignment of the system.  Custom mounts have been developed, starting with off the shelf components, which achieve our needed stability. The development and testing of these custom mounts is described in detail in Kautz et al., in these proceedings\cite{maggie_mount}.

\subsection{Pyramid WFS}

The high order WFS (HOWFS) of MagAO-X is a modulated pyramid WFS, based largely on the LBT/MagAO design.  Some subtle differences are that it is an all-reflective design, and the modulator is exactly in a pupil plane in collimated space.  The pyramid itself is an exact copy of the double 4-sided pyramid prism in existing MagAO.  The detector used is an OCAM-2K EMCCD from First Light Imaging.  A detailed design study was carried out to optimize the placement and size of the 4 pupil images on the OCAM-2K in bin-2 mode, allowing us to run the system at speeds up to 3630 Hz.  For a full description of this design and initial results from the PWFS in the lab see Schatz et al. in these proceedings\cite{lauren_pwfs}.

\subsection{Coronagraphs}

Our initial configuration is based on the vector Apodizing Phase Plate (vAPP) coronagraph.  APP coronagraphs are pupil-plane only devices, utilizing a phase pattern in a pupil plane to change the distribution of intensity in the focal plane.  See Snik et al. in these proceedings for a review of vAPP coronagraphy\cite{frans_vapp}.  We have demonstrated vAPPs in the Clio camera on existing MagAO\cite{2017ApJ...834..175O}.  A prototype MagAO-X version has been tested in our lab\cite{kelsey_lowfs}, and the final design is nearly complete and will be fabricated in the coming months.  

MagAO-X is designed to accommodate any Lyot-type coronagraph.  In our final configuration we plan to implement the Phase Induced Amplitude Apodization Complex Mask Coronagraph (PIAACMC)\cite{2010ApJS..190..220G}.  The preliminary design is underway, and we are investigating strategies for fabrication and the required tolerance of the focal plane complex mask zones.  For more information about the steps being taken to perfect the fabrication and modeling of complex focal plane masks see Knight et al. in these proceedings\cite{justin_piaacmc}.

\section{OPTICAL SPECIFICATIONS}

The specifications for the optics of MagAO-X are set to control the static and non-common path (NCP) wavefront errors (WFEs).  These sources of WFE affect the Strehl ratio and the post-coronagraph contrast.  The requirements on the optical surface quality are derived from the high-level science requirements, in conjunction with the performance simulations and error budget.  They are 
\begin{itemize}
\item Requirement: Static/NCP WFE $<$ 45 nm rms.  At H$\alpha$ this multiplies Strehl ratio by 0.83.  This specification ensures that the bright star Strehl ratio requirement of 70\% can be met.
\item Goal: Static/NCP WFE $<$ 35 nm rms. Strehl ratio ($S$) due to static/NCP impacts exposure time as $t$$\propto$$S^3$.  Hence we should seek maximum possible performance.
\item Requirement: Static/NCP WFE contribution to contrast should be $<$ $1\times10^{-3}$ from 0 to 24 $\lambda/D$.  This is after the effects of low-order WFS (LOWFS) and/or focal-plane WFS (FPWFS) are taken into account.
\end{itemize}

All of these requirements are derived for the science channel.  The WFS channel requirement is simply that it should not be worse, as it has fewer optics.  Here we briefly summarize the analysis.  Please refer to the MagAO-X PDR documents (\url{https://magao-x.org/docs/pdr}, section 5.1) for an in-depth discussion.

\subsection{The Clay Telescope}

MagAO-X is designed for the 6.5 m Magellan Clay telescope.  We collected the as-built measurements of the surfaces of the primary mirror (M1), the f/11 fixed secondary (M2), and the tertiary mirror (M3).

\noindent{\bf M1}: The Magellan Clay primary mirror was cast and polished at the Steward Observatory Mirror Lab (SOML).  From the post-polishing test data we found that the surface height has 12.52 nm rms across all spatial frequencies sampled by the map.  The central obscuration is 29\% (defined by the secondary baffle) and we will undersize the outer edge by 4.4\% to mask the poorly polished edge. This annulus has a 12.52 nm rms, only negligibly reduced from the unmasked map.

\noindent{\bf M2}: A post-fabrication measurement of the surface structure function of the Clay f/11 static secondary was available.  We fit this to a power law with index $\alpha-2$, where $\alpha$ is the index of the spatial PSD.  The measurements indicate that the surface of M2 has a residual of 12.7 nm rms.

\noindent{\bf M3}: The manufacturer's test report for the (now re-coated) tertiary mirror contained a surface map.  From this, we find that  M3 has a reported surface finish of 13.8 nm rms.

\subsection{Deformable Mirrors}

Here we describe the specifications of the deformable mirrors we will use for wavefront correction.  

\subsubsection{High Order DM}

\begin{figure}[t!]
\centering
\includegraphics[width=6in]{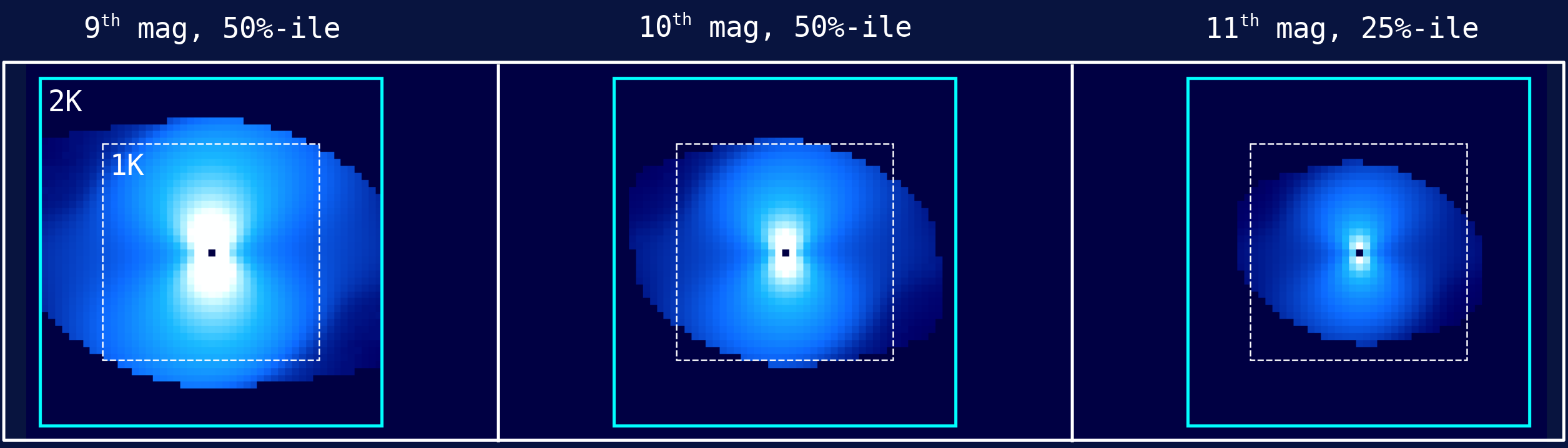}
\caption{\label{fig:2kmodeopt} Optimum gain maps for the Fourier basis, using the framework developed in Ref. \citenum{2018JATIS...4a9001M}.  The size of the modal basis in spatial frequency space is indicated for 1K and 2K DMs.  This demonstrates that the choice of a 2K DM is well justified by our requirements.}
\end{figure}

The high order DM, or tweeter, is a Boston Micromachines Corp. (BMC) MEMS 2k.  This is a 2040 actuator device, with 50 actuators across a circular aperture.    The main optical characteristics are summarized in Table \ref{tab:bmc}.  In Figure \ref{fig:2kmodeopt} we show gain maps for the Fourier basis, optimized using the framework developed in Ref. \citenum{2018JATIS...4a9001M}.  The size of the modal basis in spatial frequency space is indicated for 1K and 2K DMs, demonstrating that the choice of a 2K DM is well justified by our science case and performance requirements.

\begin{table}
\centering
\caption{Boston Micromachines 2k Specifications. \label{tab:bmc}}
\begin{tabular}{lcccccccc}
       & Total & Linear &  Pitch  & Stroke & CA   & \multicolumn{3}{c}{Flat surface [nm rms]}  \\
       & Act.  &  Act.  & [mm]    &  $\mu$m &mm    &  Max Spec       & Typical               & Actual    \\
\hline
Value: & 2040  & 50     & 0.4     & 3.5    &   19.7 &    20           &  13  & 11.3  \\
\end{tabular}
\end{table}

Delivery of the MagAO-X DM is expected in July 2018.  The device itself has been wired, coated, tested, and is undergoing final packaging.  The chosen device has one major defect, a so-called ``bump'' which limits the stroke of the surface in one position affecting 4 actuators (which are all functional).  This ``bump'' will be partially masked by a secondary support spider, and in the coronagraphs will be covered with a circle.  Additionally there is one pair of coupled actuators, and two more minor ``bumps'' which are negligible in normal operations.  

\begin{figure}
\centering
\includegraphics[width=6in]{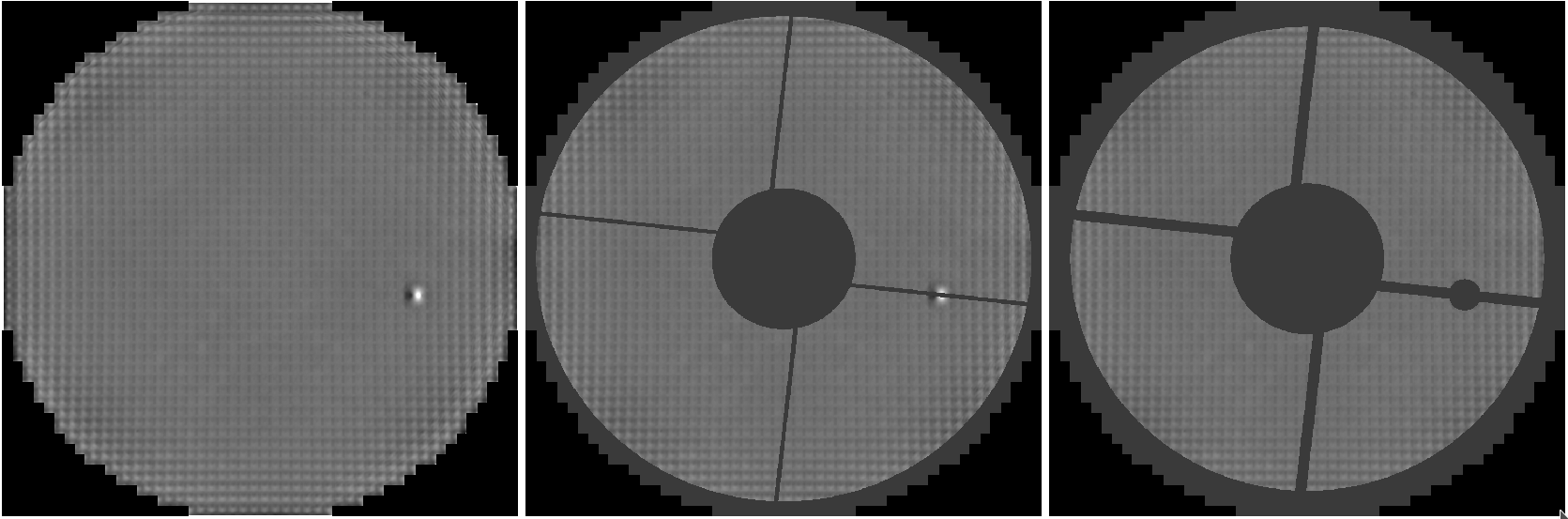}
\caption{\label{fig:2kmap} Surface maps of the 2040 actuator MEMS deformable mirror selected for MagAO-X.  The flat was measured at BMC after a closed-loop flattening procedure, and we then fit and subtracted the first 9 Zernike polynomials to simulate the woofer.  Left shows the complete surface, where the ``bump'' defect is clearly visible.  The middle pane shows the surface masked by the spiders.  At right is the coronagraphic pupil, with an oversized central obscuration, undersized outer diameter (to account for polishing errors at the edge), and spiders widened for alignment tolerances.  The masked surface is 11.3 nm rms.}
\end{figure}

BMC produced a flat map of the device.  We then subtracted the first 9 Zernike polynomials from it and analyzed the result.  With the ``bump'' masked and using our undersized coronagraphic pupil the DM flat surface has 11.3 nm rms residual.  See Figure \ref{fig:2kmap}.  Once the device is delivered we will perform an extensive characterization\cite{kyle_dms}.

\subsubsection{Low Order DM}

We plan to use an Alpao DM97-15 for our low order DMs in two positions.  The first serves as the woofer to minimize the stroke requirements on the BMC 2k tweeter.  The second will be used in the science channel (after the HOWFS beamsplitter), under control by the LOWFS.  This allows for correction of non-common path errors sensed by the LOWFS.  The main optical characteristics are summarized in Table \ref{tab:alpao}.

\begin{table}[h!]
\centering
\caption{Alpao DM97-15 Specifications. \label{tab:alpao}}
\begin{tabular}{lcccccc}
       & Total & Linear &  Pitch  & CA   & {Flat surface [nm rms]}  \\
       & Act.  &  Act.  & [mm]    & mm   &  Max Spec       & Typical                   \\
\hline
Value: & 97  & 11       & 1.5     & 13.5 &    7           &  4  \\
\end{tabular}
\end{table}

We have procured one of these devices and begun testing and characterization.  For details see Van Gorkom et al.\cite{kyle_dms} in these proceedings.

\subsection{MagAO-X Flats and OAPs}

All flats have been delivered and are undergoing final testing before integration.  These optics are better than $\lambda/40$ PV in the lowest orders.  Importantly, these are super-polished surfaces, so that at higher spatial frequencies and over the typical beam footprints in the system they perform equivalent to $\lambda/100$ PV surface or better.  Surface roughness is better than 1 Angstrom rms over the ISO-10110-8 band.  

The custom off-axis parabolas (OAPs) were also recently delivered and are being checked before integration.  Two have been used to obtain initial WFS data\cite{lauren_pwfs} and demonstrated excellent PSF image quality\cite{laird_optomech}.  These are polished to better than $\lambda/50$ rms reflected wavefront over the control band of MagAO-X, and have a surface roughness of better than 1 nm rms.  The surfaces were specified using a PSD to control WFE in bands of interest due to their effect on contrast.  See \url{https://magao-x.org/docs/pdr} section 5.1 for more detail.

\subsection{Optics Error Budget}

We have conducted a detailed analysis of the MagAO-X design with the optical specifications just discussed, to ensure that the we will meet the requirements imposed by the science case.  We analyzed each optic using a von Karman PSD with parameters characteristic of such optics.  The effects of wavefront control were modeled assuming near-perfect correction of static WFEs from optical surfaces in the control band of the system.  We categorize the resultant static and NCP errors as either uncorrectable high spatial frequency common path (i.e. ``fitting error'' of the DMs) or the post-LOWFS NCP errors.  Together these make up the instrumental WFE budget, that is not including residual atmospheric turbulence.  In-house verification of the delivered optics is ongoing.  The instrument WFE budget based on manufacturer reports is summarized in Table~\ref{tab:wfe_budget}.

\begin{table}
\centering
\caption{Instrumental WFE Budget based on as-reported fabricated optics.  Preliminary --- verification after delivery is on-going. \label{tab:wfe_budget} }
\begin{tabular}{l|c}
 & Total WFE\\
Category         & [nm rms]\\
\hline
\hline
Uncorrectable CP &  $\leq 35 nm$ \\
Post-LOWFS NCP   &  $\leq 20 nm$ \\
\hline
Total Instrumental & $\leq 40 nm$\\
\hline
\multicolumn{2}{r}{Instrumental Strehl @ H$\alpha$ = 86\%}\\
\end{tabular}
\end{table}

\subsection{Fresnel Analysis}

The above analysis only considered phase, treating the propagation as if such phase errors add in quadrature.  For the purposes of calculating total WFE and Strehl ratio this is sufficient, however it does not address the impact of various errors on contrast.  A more detailed Fresnel propagation analysis is needed to test whether the PSDs of the optical surfaces allow the system to meet the contrast requirements. For details of this analysis see Lumbres et al. in these proceedings\cite{jhen_fresnel}, where we find that as-designed MagAO-X will deliver  $6\times10^{-5}$ raw contrast at H$\alpha$ using the vAPP coronagraph.

\subsection{System Throughput}
We analyzed the throughput of the as-designed system. This ensures that we can adequately analyze and simulate system performance. 
The key quantity is the photon rate (photons/sec) delivered to each of the detector planes in the system.  These are the high-order wavefront sensor  (HOWFS), the low-order WFS (LOWFS), and the two SDI science cameras.  

The following coatings were assumed:
\begin{itemize}
\item Primary and Secondary: protected Aluminum
\item Tertiary: a custom coating with enhanced reflectivity near 0.8 $\mu$m.
\item OAPs and Flats: protected Silver, using a curve provided by Thor Labs.
\item Alpao DMs: same protected Silver curve.
\item BMC MEMS DM: unprotected gold
\item Transmissive surfaces: a standard anti-reflective coating.
\end{itemize}

The quantum efficiency (QE) curves for the First Light Imaging OCAM-2K (HOWFS detector) and the Andor iXon 897 (LOWFS detector) and Princeton Instruments EMCCDs (Science detectors) were digitized from the manufacturer specification sheets.  We include atmospheric transmission calculated using the BTRAM IDL code.  We assumed 5.0 mm precipitable water vapor (PWV), and observing at zenith distance 30$^o$, or airmass 1.15.

We then created a notional set of filters designed for the H$\alpha$ science case:
\begin{itemize}
\item A dichroic beamsplitter which divides the light between the HOWFS and Science channels.  Cuts-on at 0.68 $\mu$m, with $T$ and $R$ both 95\%.
\item A LOWFS filter which selects only in-band light of the vAPP leakage term, including the 5\% leakage term.
\item An H$\alpha$ SDI filter set, as quoted by a vendor (see Opto-Mechanical design for details).
\end{itemize}

\begin{figure}[h!]
\centering
\includegraphics[width=4in]{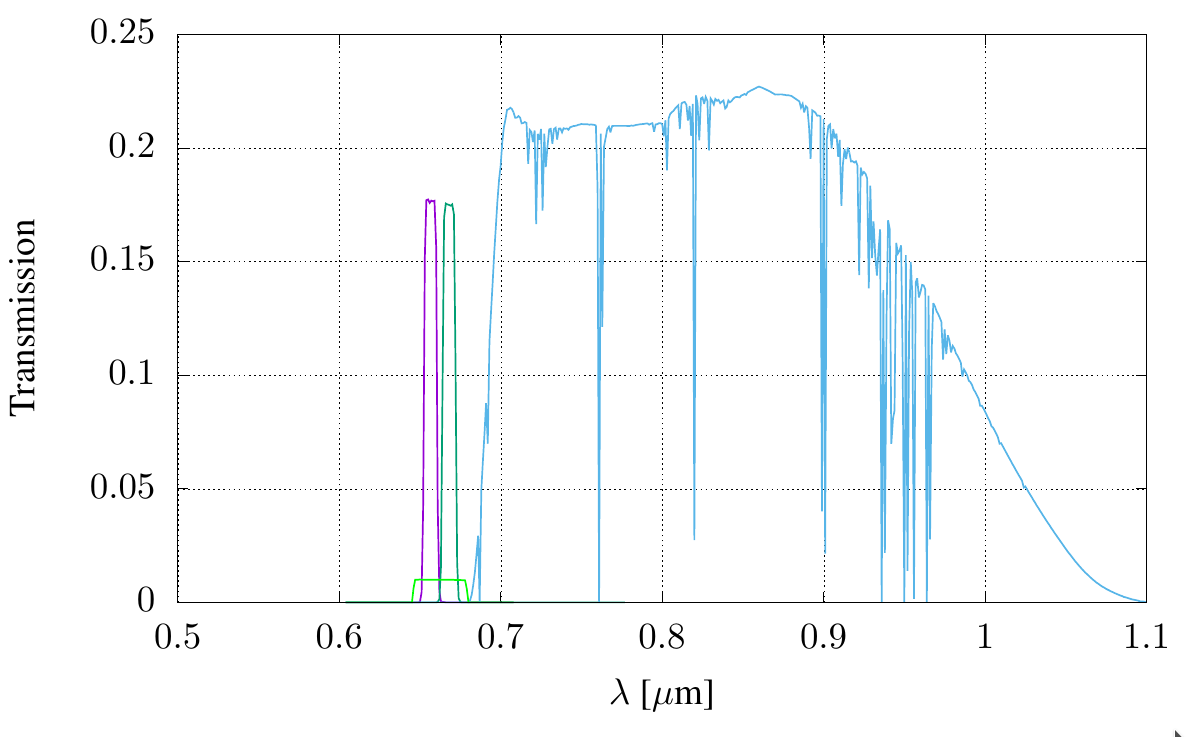}
\caption{Filter transmission curves for MagAO-X at each of the detector planes for H$\alpha$ SDI. Blue: PWFS,  green: LOWFS, maroon \& dark green: SDI cameras.  \label{fig:trans_final}}
\end{figure}

Finally, the complete transmission curves for each of the planes in the H$\alpha$ configuration was calculated by multiplying the above curves.  This the reflectance or transmittance for each optic in the system, the atmosphere, the detector QE, and the appropriate filter curves.  We also included a 10\% loss due to diffraction derived from the Fresnel propagation analysis.  These final transmission curves are shown in Figure \ref{fig:trans_final}.  The characteristics of this filter system are given in Table~\ref{tab:char_halpha}.

\begin{table}[h!]
\centering
\caption{Filter characteristics for the H$\alpha$ configuration. \label{tab:char_halpha}}
\begin{tabular}{lccccl}
Plane  &  Throughput  &  $\lambda_0$   &  $\Delta \lambda$  &  $F_\gamma$(0) &  Notes\\
       &              &    $\mu$m      &    $\mu$m          &  Photons/sec & \\
\hline
Science H$\alpha$ & 0.177 & 0.657 & 0.0082 & $2.3\times 10^{9}$ & \\
Science Cont.     & 0.176 & 0.668 & 0.0083  & $2.6\times 10^{9}$ & \\
LOWFS             & 0.010 & 0.662 & 0.033 & $5.9\times 10^{8}$ & 5\% vAPP leakage\\
HOWFS             & 0.227 & 0.851 & 0.257  & $7.6\times 10^{10}$ & \\
\hline
\end{tabular}
\end{table}

\section{Compute and Control}

The MagAO-X compute system hardware is entirely commercial off-the-shelf (COTS), utilizing standard components.  There are three computers: the Instrument Control Computer (ICC), the Real-Time computer (RTC), and the AO Control Computer (AOC).  The fundamentals of these computers are identical, meaning they use the same motherboard, processor, and RAM.  This facilitates sparing.  Because these are all common COTS components, it also allows for managing obsolescence by making it likely that replacements will be available for some time, and furthermore eventual upgrades should be simple to manage.

Each machine  uses dual 16 core Intel Xeon processors.  For additional computing power we employ COTS GPUs, for now relying on consumer grade GPUs rather than the much more expensive units intended for scientific and engineering use.  Experience on SCExAO has shown this is adequate for MagAO-X closed loop operation at 3.63 kHz.  The mixed CPU/GPU architecture can be scaled up to meet additional computing requirements required to support advanced AO operation modes, such as predictive control\cite{2018JATIS...4a9001M,2017arXiv170700570G} and multi-sensor operation\cite{2017arXiv170700570G}.

Our real time software is based on the ``Compute and Control for Adaptive Optics'' (CACAO) system, developed for SCExAO and now being refactored significantly to support broad usage in ExAO systems.  See \url{https://github.com/cacao-org/cacao} for more information about CACAO.

\subsection{Telemetry}

An important challenge in high-order and high-speed ExAO systems is management of telemetry.  We estimate that, when fully operational, MagAO-X could generate as much as 10 TB of data per night.  A key feature of our computing system is that it is designed to store all system telemetry all the time, and we plan to save this to facilitate on-line optimization, experiments in post-processing, and off-line system analysis.  This is manageable on-site with COTS hardware, but this volume of data will require physical transport from LCO back to UofA.  See \url{https://magao-x.org/docs/pdr} Section 3.2 for details.

Our motivations for recording system-wide telemetry for later use include:
\begin{itemize}
\item Continuous updates (on timescales of $\sim$ 1 minute) of machine learning predictive controllers\cite{2017arXiv170700570G}
\item Analysis of system optimization
\item Post-processing estimation of PSF residuals to assist PSF subtraction/calibration
\end{itemize}

Using such large volumes of information in post-processing will be very challenging.  We have been working to develop computational systems capable of handling such large data sets\cite{2016SPIE.9913E..0FH}.  When employing PSF estimation strategies such as KLIP, matrices with sizes set by the number of images must be decomposed.  Long et al. in these proceedings\cite{joseph_sirius} describe work to use vAPP observations with existing MagAO\cite{2017ApJ...834..175O} to develop a pipeline capable of reducing $>10^4$ images at a time with advanced dimensionality reducing algorithms.   Future work will explore using system telemetry, in addition to focal plane images, to improve detection limits.

\section{WAVEFRONT CONTROL}

\subsection{High Order Wavefront Control}

The high order, meaning both high spatial and high temporal frequencies, is provided by the PWFS running at speeds up to 3.63 kHz in closed loop with the BMC 2040 actuator MEMS HODM.  To ensure that the HODM is operated around its mid-bias region and to deal with aberrations outside its 3.5 $\mu$m stroke, a woofer is needed.  We considered using the ASM, but have settle on the Alpao DM-97 as the woofer.  This device can operate at up to 2 kHz.  

We have considered three strategies for managing this split control system:
\begin{enumerate}
\item Cascaded control: The existing MagAO is used as an initial cleanup stage, and MagAO-X functions as an independent ``afterburner''.  This was our original concept, and is included here for comparison.
\item Split control: The woofer and tweeter are essentially treated as a single DM for the purposes of wavefront control.  This necessitates operating the system no faster than the 2 kHz speed of the woofer.   
\item Offloading: The tweeter receives all commands directly from the PWFS at full rate, but at some interval the average shape is offloaded to the woofer.  This has been analyzed and found to be stable\cite{2006SPIE.6306E..0BB}, and the BMC 2K and Alpao DM-97 have sufficient stroke to support it.
\end{enumerate}

\begin{figure}
\centering
\includegraphics[width=3in]{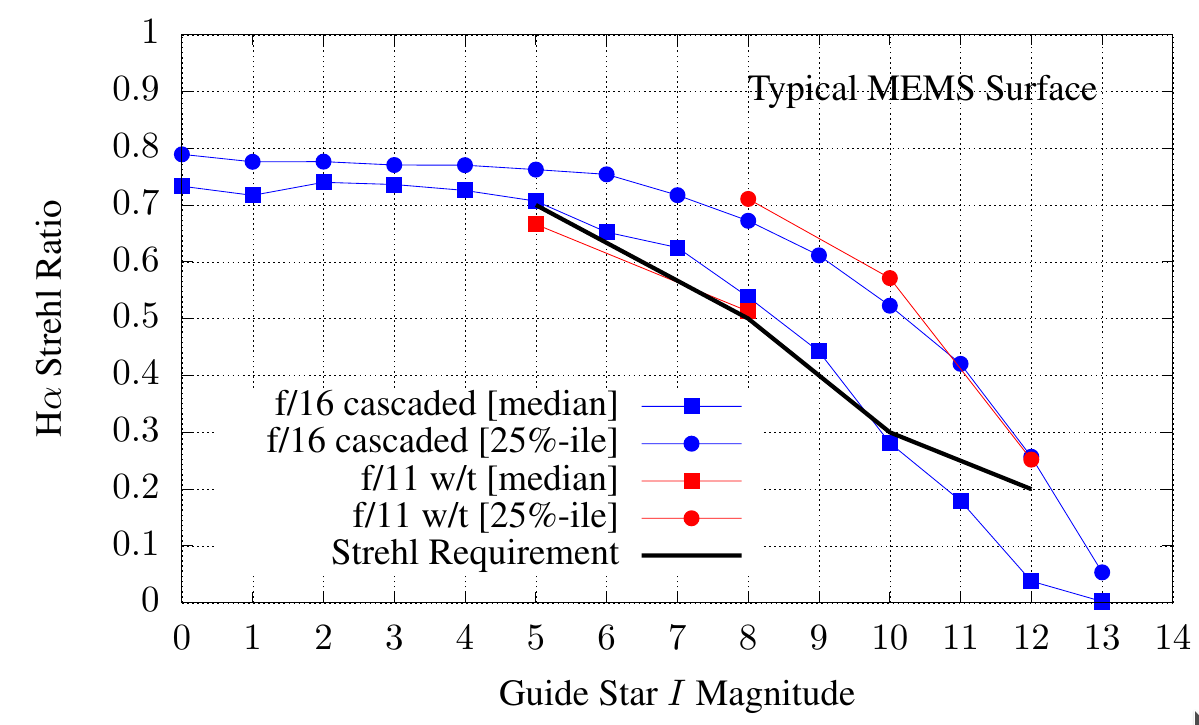}
\caption{\label{fig:sim_strehl} Simulated Strehl}
\end{figure}

The first option, cascaded control, is by far the simplest to implement and is in use on SCExAO\cite{julien_scexao}.  We show results from a suite of simulations as the blue curves in Fig \ref{fig:sim_strehl}.  However, as we discuss above this option is operationally very complex due to the requirement to have the adaptive secondary on the telescope.

The second option is a conventional method for woofer-tweeter control.  The main drawback here is the requirement to run the system at no more than 2 kHz due to the dynamic response of the Alpao DM-97.  Simulations with this method are shown as the red curves in Fig. \ref{fig:sim_strehl}.

The results in Fig \ref{fig:sim_strehl} show that the split woofer-tweeter control, even though limited to 2 kHz, will meet our requirements.  This is true in 25 percentile conditions at LCO for all guide stars, and is true for most stars in median conditions.  We have not yet implemented offloading in simulation, but because correction will occur at 3.63 kHz on bright stars we expect to exceed the requirement with the offloading control scheme.

\begin{figure}
\includegraphics[width=3in]{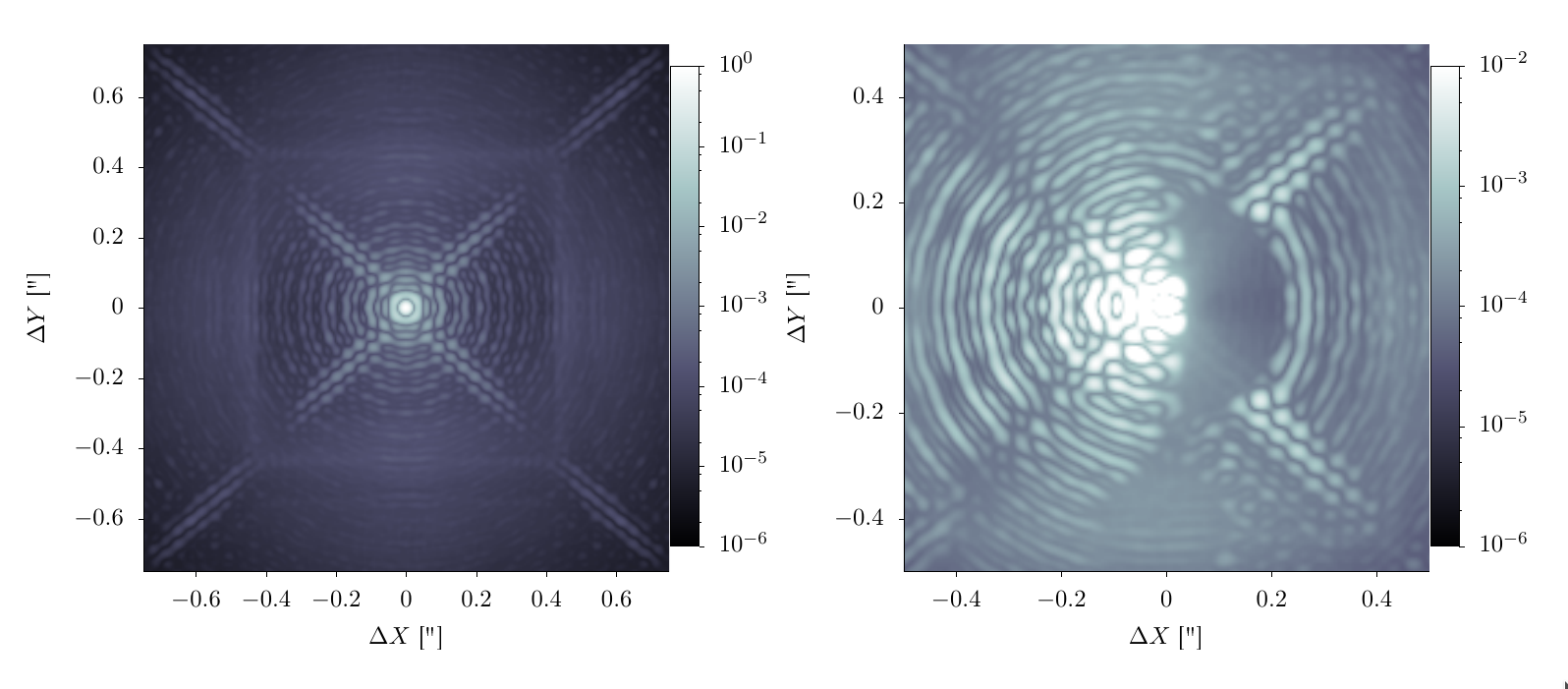}
\includegraphics[width=3in]{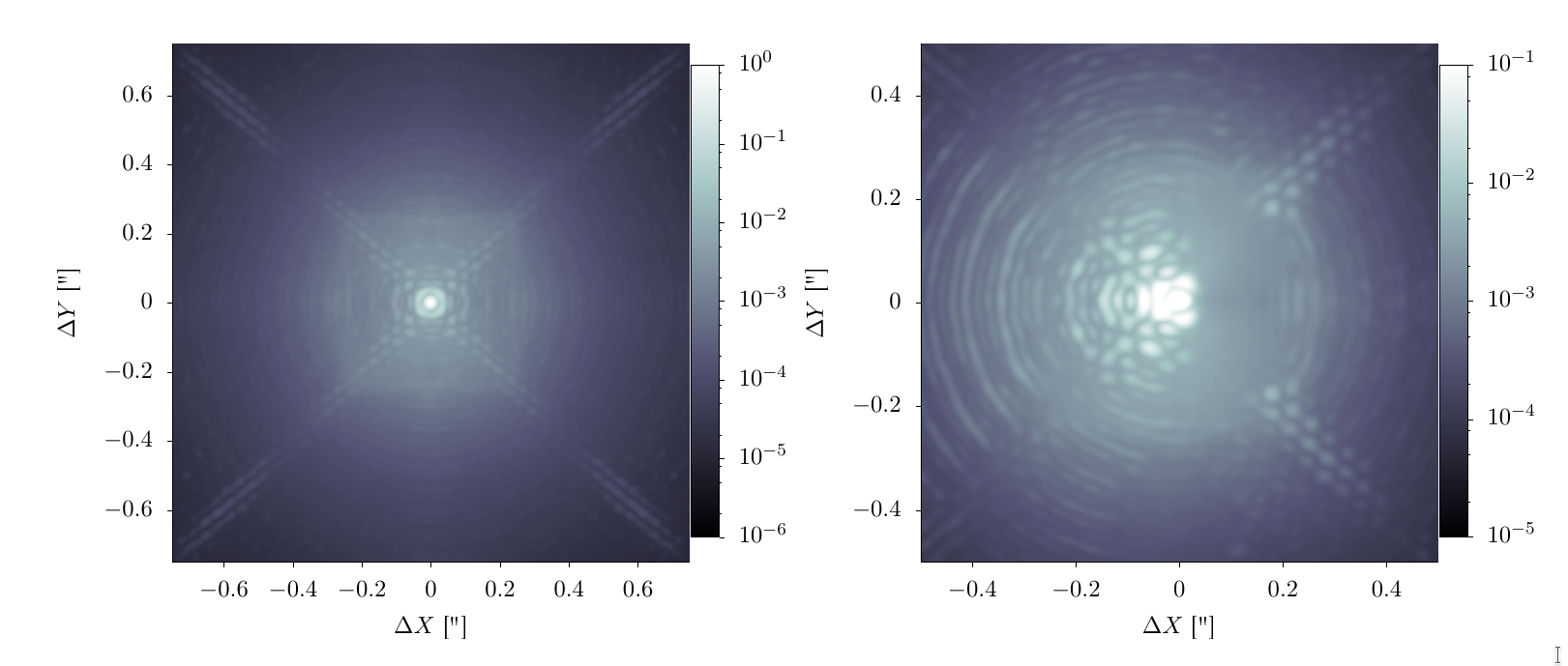}
\caption{\label{fig:sim_psfs} Simulated PSFs and post-coronagraph science images.  Left Two Images: a 5th mag guide star in median conditions.  Right Two Images: a 12th mag guide star in 25th percentile conditions.}
\end{figure}

Figure \ref{fig:sim_psfs} shows example PSFs and post-coronagraph science images for the vAPP coronagraph.  See \url{http://magao-x.org/docs/pdr} Section 4.1 for more detailed discussion of these results.

\subsection{FP/CLOWFS}

A key capability in any high-contrast imaging instrument is control of non-common-path (NCP) aberrations.  These aberrations are not sensed by the PWFS, but do affect contrast in the final science focal planes.  We are implementing several strategies for sensing and controlling these.

CLOWFS uses light rejected by the coronagraph to sense low order aberrations.  MagAO-X is designed to feed light rejected from either the focal plane mask (FPM) or the Lyot mask of the coronagraph to an EMCCD camera.  We plan to use already developed techniques\cite{2017PASP..129i5002S} for applying this in Lyot-type coronagraphs.

We are also developing several novel strategies for CLOWFS with the vAPP coronagraph.  We plan to implement an FPM at an intermediate focal plane after the vAPP which reflects all light from the image except that in the dark hole and leakage terms.  This light is then sent to the LOWFS camera.  Standard CLOWFS strategies can be applied (i.e. using out of focus images).  We are also testing the use of modal wavefront sensing spots encoded in the vAPP liquid crystals.  The amount of light in these spots relates directly to the amplitude of the corresponding aberration.  We also plan to use linear dark field control (LDFC)\cite{2017JATIS...3d9002M} to control mid-spatial frequencies.  For a detailed description of our FP and CLOWFS strategies see Miller et al. in these proceedings\cite{kelsey_lowfs}.

The LDFC technique relies on a dark hole being created using some other technique.  In the case of the vAPP this is achieved by the coronagraph directly, though one must minimize low-order and mid-order aberrations to achieve the full potential.  For our more aggressive PIAACMC architecture, we will need to implement post-coronagraphic wavefront control techniques.  One avenue we are exploring is the use of high-rate focal plane images, combined with WFS telemetry, to infer the static and NCP aberrations.  This was proposed by Frazin\cite{2013ApJ...767...21F} as a post-processing technique, and we are investigating extending it to a real-time control system.  See Rodack et al. in these proceedings\cite{alex_rtfa} for a complete introduction to this concept.

\section{CONCLUSION}

The MagAO-X project is now undergoing integration at Steward Observatory.  The design and component specifications have been carefully developed to ensure that we meet the very demanding performance requirements based on high-contrast imaging of exoplanets at visible wavelengths. Assuming remaining procurement completes in short order, we expect to be on-sky for initial testing at the Magellan Clay telescope in the 2019A semester.

In addition to the key science goal of surveying the population of H$\alpha$ emitting accreting protoplanets, we view MagAO-X as a development platform for future ExAO systems on the next generation of Giant Segmented Mirror Telescopes (GSMTs).  These large telescopes have great potential to revolutionize our knowledge of the universe.  In particular, they will be able to detect and characterize planets in much larger numbers than currently possible.  See Males et al.\cite{jared_gsmts}  and Fitzgerald et al.\cite{fitz_psi}, in these proceedings.  A key goal of this research is to develop the ability for GSMTs to characterize potentially habitable planets around nearby late-type stars\cite{2012SPIE.8447E..1XG,2014SPIE.9148E..20M}.  Though our main priority is the science cases outlined in Section 2, MagAO-X will be an enabling platform for developing, testing, and proving technologies for the GSMT ExAO systems to come.

\acknowledgments 
 
We are very grateful for support from the  NSF MRI Award \#1625441 (MagAO-X).\

\bibliography{report} 
\bibliographystyle{spiebib} 

\end{document}